\documentclass[letterpaper,11pt]{article}
\usepackage{jheppub} 

\usepackage{graphicx}

\usepackage{amsmath}
\usepackage{booktabs}

\usepackage{slashed,xcolor}
\usepackage{multirow,array,rotating}
\usepackage{morefloats}
\usepackage{color}

\def\End{\end{document}}

\def\ba{\begin{array}}
\def\ea{\end{array}}
\def\beqn{\begin{eqnarray}}
\def\eeqn{\end{eqnarray}}
\def\beqs{\begin{subequations}}
\def\eeqs{\end{subequations}}

\newcommand{\Dfb}{\mathord{\buildrel{\lower3pt\hbox{$\scriptscriptstyle{\leftrightarrow \tiny{ \ \ \ } }$}}\over {D^{\mu}}}} 

\newcommand{\Dfbd}{\mathord{\buildrel{\lower3pt\hbox{$\scriptscriptstyle\leftrightarrow$}}\over {D}_{\mu}}}

\newcommand{\gsim}{\buildrel > \over {_\sim}}
\newcommand{\met}{\mbox{${\not\! E}_{\rm T}$}}

\def\h{h^0}
\def\H{H^0}
\def\A{A}
\newcommand{\hc}{H^{\pm}}

\title{\bf The Higgs Portal and Cosmology}

\author[a]{Ketevi Assamagan,}
\author[b,c]{Chien-Yi Chen,}
\author[d]{John Paul Chou,}
\author[e]{David Curtin,}
\author[f]{Michael A. Fedderke,}
\author[d]{Yuri Gershtein,}
\author[g]{Xiao-Gang He,}
\author[h]{Markus Klute,}
\author[i]{Jonathan Kozaczuk,}
\author[j]{Ashutosh Kotwal,}
\author[k]{Steven Lowette,}
\author[l]{Jose Miguel No,}
\author[m]{Tilman Plehn,} 
\author[n]{Jianming Qian,}
\author[o]{Michael Ramsey-Musolf,}
\author[p]{Alexei Safonov,} 
\author[q]{Jessie Shelton,}
\author[r]{Michael Spannowsky,}
\author[s]{Shufang Su,}
\author[t]{Devin G. E. Walker,}
\author[o]{Stephane Willocq,}
\author[o]{Peter Winslow}
\affiliation[a]{Department of Physics, Brookhaven National Laboratory, Upton, New York, 11973}
\affiliation[b]{Perimeter Institute for Theoretical Physics, Waterloo, ON N2L 2Y5, Canada}
\affiliation[c]{Department of Physics and Astronomy, University of Victoria, Victoria, BC V8P 5C2, Canada}
\affiliation[d]{Department of Physics and Astronomy, Rutgers University, 126 Frelinghuysen Road. Piscataway, NJ 08854-8019}
\affiliation[b]{Maryland Center for Fundamental Physics, Department of Physics,
University of Maryland, College Park, MD 20742-4111 USA}
\affiliation[f]{Department of Physics and Kavli Institute for Cosmological Physics, The University of Chicago, Chicago, IL 60637, USA}
\affiliation[g]{INPAC, SKLPPC, and Department of Physics, Shanghai Jiao Tong University, Shanghai, China}
\affiliation[h]{Department of Physics, Massachusetts Institute of Technology, Cambridge, MA 02139 USA}
\affiliation[i]{TRIUMF, 4004 Wesbrook Mall, Vancouver BC V6T 2A3, Canada.}
\affiliation[j]{Department of Physics, Duke University,Physics Bldg., Science Dr., Durham, NC 27708 and Fermilab, Batavia IL 60510-5011}
\affiliation[k]{Physics Department, Vrije Universiteit Brussel, Pleinlaan 2 | 1050 Brussel, Belgium}
\affiliation[l]{Department of Physics and Astronomy, University of Sussex,  Brighton,  BN1 9QH, UK}
\affiliation[m] {Institute for Theoretical Physics,Heidelberg University,Philosophenweg 16, 69120 Heidelberg, Germany }
\affiliation[n] {Department of Physics, Randall Laboratory, University of Michigan,
450 Church Street, Ann Arbor, MI  48109-1040}
\affiliation[o] {Amherst Center for Fundamental Interactions, Department of Physics, University of Massachusetts Amherst, Amherst, MA 01003 USA}
\affiliation[p] {Department of Physics and Astronomy, Texas A \& M University, 
College Station, TX 77843-4242}
\affiliation[q]{Department of Physics, University if Illinois Urbana-Champaign, Urbana, IL 61801-3080}
\affiliation[r]{Department of Physics, Durham University, South Road, 
Durham, DH1 3LE, UK}
\affiliation[s] {Department of Physics, University of Arizona, Tucson, AZ 85721}
\affiliation[t] {Department of Physics, University of Washington, Seattle, WA 98195, USA}

\abstract{
Higgs portal interactions provide a simple mechanism for addressing two open problems in cosmology: dark matter and the baryon asymmetry. In the latter instance, Higgs portal interactions may contain the ingredients  for a strong first order electroweak phase transition as well as new CP-violating interactions as needed for electroweak baryogenesis. These interactions may also allow for a viable dark matter candidate. We survey the opportunities for probing the Higgs portal as it relates to these questions in cosmology at the LHC and possible future colliders.
}

\begin{document}
\maketitle
\baselineskip 18.5pt


\section{Introduction}
\label{sec:intro}

Explaining the origin and composition of the matter content of the Universe remains one of the most compelling tasks at the interface of high energy physics, nuclear physics, and cosmology. The identity of the dark matter that comprises 27\% of the cosmic energy density remains undetermined, and little is known about its non-gravitational interactions. The visible matter comprises just under 5\% of the present cosmic energy density and is often characterized by the baryon-to-photon ratio 
\begin{eqnarray}
Y_B=\frac{n_B}{s} = (8.59 \pm 0.11) \times 10^{-11} 
\end{eqnarray}
where $n_B$ ($s$) is the baryon number (entropy) density and where the value has been obtained from the Planck data\cite{Ade:2013zuv}. This number, though tiny, is clearly decisive for the Universe as we know it, yet the Standard Model (SM) suggests it should be many orders of magnitude smaller. Thus, accounting for the abundance of both the visible and dark matter
provides some of the strongest motivation for physics beyond the Standard Model (BSM). 

It is possible that the dynamics associated with the dark matter and the origin of the baryon asymmetry are largely hidden from our view, either because the associated mass scale is too high or the relevant interactions with SM particles too feeble. In light of the discovery of the Higgs-like boson at the LHC, it is interesting to ask whether the properties and interactions of the Higgs boson provide a window, or \lq\lq portal", on the origin and composition of the cosmic matter content. If so, what might one learn from more refined studies of Higgs boson properties and interactions at the LHC or from the search for additional Higgs-like states? The purpose of this document is to summarize the landscape of possibilities as reviewed at the workshop \lq\lq Unlocking the Higgs Portal" held at the Amherst Center for Fundamental Interactions at the University of Massachusetts in May 2014\cite{ACFI14}. In view of  Run II of the LHC, it is particularly worthwhile to identify the Higgs boson properties and searches for new states that are most promising from the standpoint of the cosmic matter content problem. In what follows, we provide a snapshot of this landscape as well as a discussion of additional theoretical work needed to delineate the prospective consequences of future LHC studies. Given the prospects for developing the next generation of high energy colliders, we also discuss the prospective opportunities for probing cosmologically relevant Higgs portal scenarios at various new facilities under consideration, including the International Linear Collider (ILC), China Electron Positron Collider (CEPC) and Super Proton Proton Collider (SppC), and the CERN future circular colliders FCC-ee (electron-positron) and FCC-hh (proton-proton).

The collider phenomenology pertaining to dark matter has been extensively investigated, particularly in relation to the WIMP paradigm (see, {\em e.g.}, Refs.~\cite{Morrissey:2009tf,Abdallah:2015ter,Abercrombie:2015wmb}). Dark matter-Higgs boson interactions have also been widely studied, and the high energy community is reasonably well versed in the dark matter problem (for recent discussions, see, {\em e.g.}. Refs.~\cite{Abdallah:2014hon,Craig:2014lda}).  The implications of present and future LHC studies for the origin of visible matter, on the other hand, is less widely appreciated. Consequently, in what follows we will place a somewhat heavier weight on the baryogenesis problem, linking it to the Higgs portal and dark matter relic abundance where appropriate. For completeness, we also provide a short summary of the collider phenomenology of Higgs portal dark matter. 

In principle, the non-vanishing $Y_B$ could have resulted from initial conditions during the Big Bang or from dynamics of grand unified theories at energy scales above $\sim 10^{16}$ GeV. In practice, the on-going success of the inflationary paradigm implies that any matter-antimatter asymmetry created in either of these ways would have been inflated away and, thus, not able to account for present observation. In this view, the particle physics of the post-inflationary Universe (including the era of preheating) is likely responsible. Over 40 years ago,  Sakharov\cite{Sakharov:1967dj} identified three ingredients in the early Universe that must have been present in order to generate a non-vanishing $Y_B$: (a) baryon number (B)-violation; (b) violation of both C- and CP-invariance; and (c) either a departure from equilibrium dynamics or violation of CPT-invariance. The SM contains the first ingredient in the form of (B+L)-violating sphaleron transitions, but fails on the second and third. 

The possible BSM scenarios that satisfy all three \lq\lq Sakharov" conditions span the gamut of post-inflationary cosmic history. Among the most theoretically attractive and experimentally testable are those that introduce new particles in the few-hundred GeV to TeV mass range. These scenarios would have generated the matter-antimatter asymmetry during or shortly before the era of electroweak symmetry-breaking (EWSB). The most thoroughly studied (though not exclusive) such scenario is electroweak baryogenesis (EWBG) (for a recent review, see Ref.~\cite{Morrissey:2012db}).

In EWBG, the Universe undergoes a first order phase transition during which electroweak symmetry is broken. The electroweak phase transition (EWPT) proceeds via nucleation of bubbles of broken electroweak symmetry as the Universe cools through a nucleation temperature $T_N$ that lies below the phase transition critical temperature, $T_C$. This transition, which satisfies the Sakharov out-of-equilibrium condition,
is analogous to the condensation of water droplets from vapor with decreasing temperature. Sakharov's second ingredient is provided by C- and CP-violating interactions of new particles at the bubble walls. These interactions ultimately induce the sphalerons to create baryons that diffuse inside the expanding bubbles where they are captured and protected from being washed out by inverse sphaleron processes.

The LHC  and prospective future colliders are well-suited to looking for the particle physics ingredients needed for the first order EWPT. Indeed, the possibilities for generating this transition are rich. New particles may modify the Higgs potential through either loop effects or new tree level interactions. In some scenarios, the result may
be new patterns of EWSB that would not arise in the Standard Model, such as the occurrence of a series of transitions that break SM symmetries. In all cases, the dynamics require the existence of new spin-zero particles whose interactions may be more or less analogous to those of the SM Higgs boson. If one (or more) of these new states is stable on cosmological time scales, it (they) may also account for the dark matter relic density. Alternatively, dark matter fields may interact with the new scalars that, in turn, interact with the SM through the Higgs portal. 

Generically, we write the Higgs portal interaction as
\beqn
\label{eq:portal1}
\mathcal{L} \supset \frac{a_1}{2} H^\dag \phi H +\frac{a_2}{2} H^\dag H \phi^\dag\phi+\cdots\ \ \ ,
\eeqn 
where $H$ is the SM Higgs doublet, $\phi$ is an additional scalar transforming as either a singlet or non-singlet under the SM, and the \lq\lq $+\cdots$" indicate possible higher dimensional operators. For the neutral component of $\phi$ to contribute to the dark matter relic density, the $Z_2$-odd terms in  Eq.~(\ref{eq:portal1}) must be absent ({\em e.g.}, $a_1\to 0$). On the other hand, $\phi$ may not itself be the dark matter candidate but may interact independently with the dark matter as in
\beqn
\label{eq:portal2}
\mathcal{L} \supset {\bar\chi}\left(a+b\gamma_5\right)\phi\chi +\mathrm{h.c.}+\cdots\ \ \ ,
\eeqn
where $\chi$ in this case is fermionic dark matter. The interactions (\ref{eq:portal1},\ref{eq:portal2}) may provide the Higgs portal into the dark matter sector, while the operators in Eq.~(\ref{eq:portal1}) may give rise for a strong first order EWPT as needed for EWBG.

The LHC and future colliders might discover the new particles ($\phi$, $\chi$) and probe the effects of their interactions in a number of ways:
\begin{itemize}
\item modified SM-like Higgs boson couplings to itself and other SM particles
\item new Higgs boson production mechanisms
\item new Higgs boson decay channels
\item new scalar particles that interact with the Higgs boson and other SM particles
\end{itemize}
Many of these signatures have been discussed elsewhere in the literature. Here we discuss their relation to the possible occurrence of a first order EWPT and connection to dark matter, utilizing representative theoretical scenarios. 



\section{Theoretical Scenarios}
\label{sec:theory}

The dynamics of the EWPT are governed by the finite-temperature effective action, $S_\mathrm{EFF}(T)$, that reduces to an integral over the effective potential $V_\mathrm{EFF}(\phi, T)$ for spatial homogenous background fields $\phi$ (here, we generically denote the full set of background fields by $\phi$). For a theory such as the SM wherein only a single field acquires a vacuum expectation value (the background field), one may write the effective potential in the high-temperature limit
\beqn
V_\mathrm{EFF}(\varphi, T) = D(T^2-T_0^2)\varphi^2 -  (E T+e) \varphi^3 + {\bar\lambda}\varphi^4 +\cdots\ \ \ ,
\label{eq:veff}
\eeqn
where $D$, $T_0^2$, $E$, $e$, and ${\bar\lambda}$ are all computable from the zero temperature Lagrangian. In the SM, where $\varphi$ is the Higgs background field (used here interchangeably with the vev), ${\bar\lambda}$ is approximately the Higgs self-coupling. The quantity $D T^2$ corresponds to the square scalar field thermal mass, while $DT_0^2$ is the zero temperature, tachyonic mass parameter, often denoted $\mu^2$. The existence of the cubic term proportional to $ET+e$ is essential for the occurrence of a first order phase transition. In the SM, $e=0$, as there exists no tree-level term that is cubic in the background field. In extended scalar sectors, $e$ may be non-vanishing. The quantity $E$ is generated at loop level in both the SM and its extensions. 

\begin{figure}[ht!]
  \centering
  \includegraphics[width=0.5\linewidth]{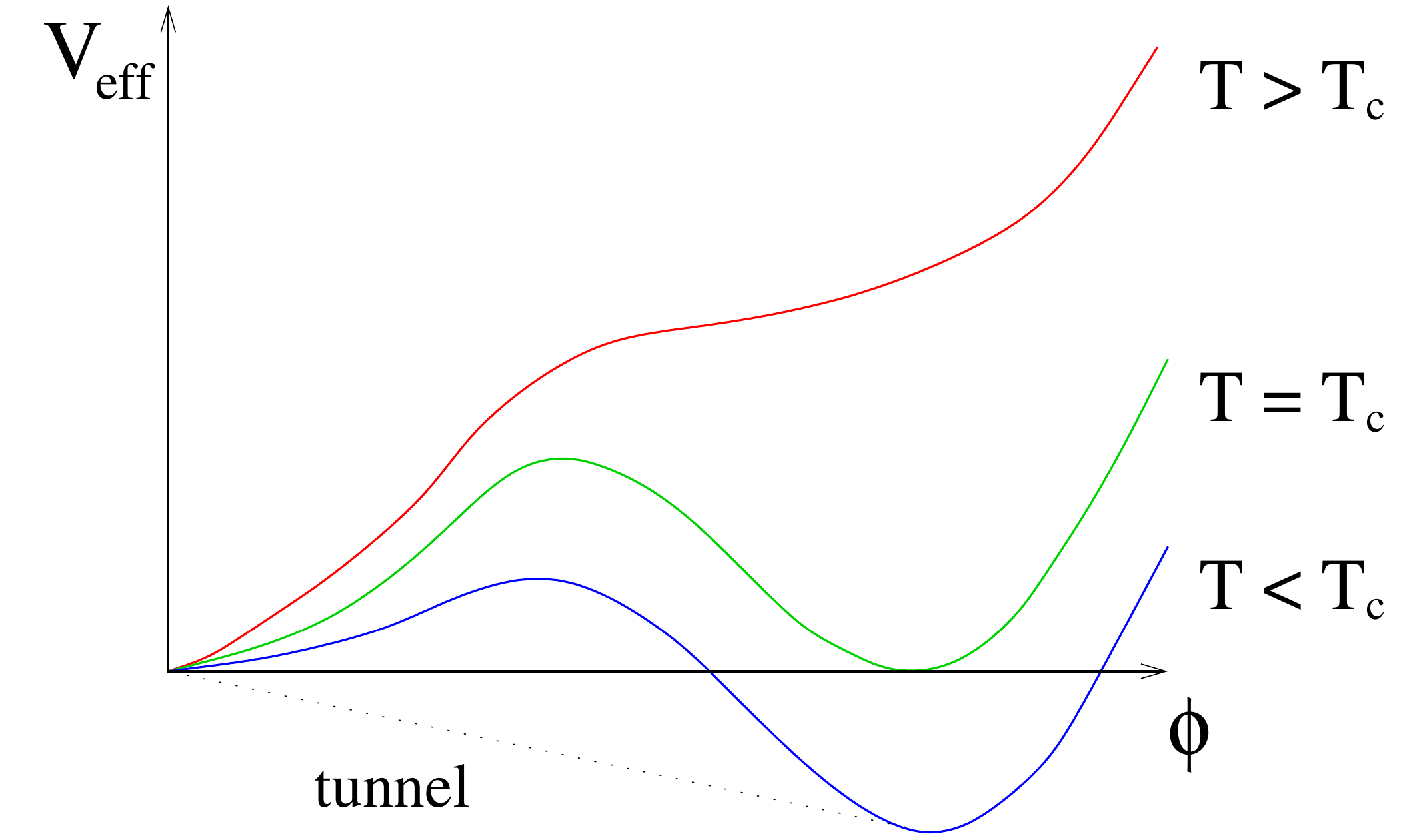}
  \caption{\label{fig:veff}Evolution of the effective potential with $T$. }
\end{figure}

The interplay between $(ET+e)$ and ${\bar\lambda}$ governs the character of the EWPT. In the limit that both $E$ and $e$ vanish, the transition becomes second order and no bubble nucleation will occur. The evolution of the potential with temperature for a first order EWPT is shown in Fig.~\ref{fig:veff}. The first order transition is marked by the existence of a barrier  between the unbroken ($\varphi=0$) and broken ($\varphi\not=0$) minima that requires $E>0$ and/or $e>0$. In a second order transition the cubic terms in the finite temperature potential do not appear. 

Assuming the occurrence of a first order transition with a sufficiently large bubble nucleation rate, preservation of any matter-antimatter asymmetry inside the bubbles requires quenching of the sphaleron transitions by making the sphaleron energy sufficiently large relative to the critical temperature, $T_C$ . For the high-$T$ potential of Eq.~(\ref{eq:veff}), the requirement becomes\cite{Quiros:1999jp,Profumo:2007wc,Patel:2011th}
\beqn
\label{eq:sfoewpt}
\frac{(ET_C + e)}{{\bar\lambda} T_C}\gsim 1\ \ \ .
\eeqn
This criterion is sometimes referred to as the requirement for a \lq\lq strong first order EWPT" or \lq\lq SFOEWPT".
In the SM, for which $e=0$ and $E$ arises through loops, this criterion is not satisfied, largely because the Higgs self coupling $\lambda$ is too large, thereby suppressing the denominator in Eq.~(\ref{eq:sfoewpt}). Since $m_h^2=\lambda v^2$, where $v$ is the $T=0$ vev, the requirement that the denominator of Eq.~(\ref{eq:sfoewpt}) be sufficiently small is equivalent to an upper bound on the Higgs boson mass. In fact, Monte Carlo studies indicate that  the maximum Higgs boson mass for a first order EWPT in the SM is in the vicinity of 80 GeV\cite{Gurtler:1997hr,Laine:1998jb,Csikor:1998eu,Aoki:1999fi}, well below the observed value. 

BSM scenarios may remedy the absence of a SFOEWPT by increasing the magnitude of E, reducing the magnitude of ${\bar\lambda}$, or reducing the critical temperature $T_C$. Broadly speaking, BSM scenarios do so {\em via} one or more of the following avenues:
\begin{itemize}
\item new loop effects that effectively increase $E$ (Section \ref{sec:loop})
\item new tree-level interactions that generate non-vanishing $e$ (Section \ref{sec:tree})
\item tree-level interactions that reduce ${\bar\lambda}$ either directly or indirectly (Section \ref{sec:tree})
\item tree-level or loop effects that lower the critical temperature, $T_C$ (Section \ref{sec:tree}).
\item new tree-level interactions that generate an earlier SFOEWPT along a different field direction, enabling EWBG through a multi-step process (Section \ref{sec:tree})
\end{itemize}

\subsection{Loop effects}
\label{sec:loop}
The dynamics of a loop-induced SFOEWPT can be understood by considering the so-called \lq\lq daisy resummation" contribution to $V_\mathrm{EFF}(\varphi, T)$. This contribution yields the following contribution to the difference in energy between the potential in the broken and unbroken phases:
\beqn
\label{eq:deltav}
\Delta V_\mathrm{daisy} = -\frac{T}{12\pi}\sum_k\left\{ \left[m_k^2+ y_k \varphi^2 + \Pi_k(T)\right]^{3/2} - \left[m_k^2+ \Pi_k(T)\right]^{3/2}\right\}\ \ \ ,
\eeqn
where the sum is over all bosonic degrees of freedom with mass parameters $m_k$, couplings to the Higgs field $y_k$, and thermal masses $\Pi_k(T)$. The presence of $\Delta V_\mathrm{daisy}$ effectively generates a barrier between the two phases. For the transverse components of the $W$ and $Z$ bosons, both $m_k^2$ and $\Pi_k(T)$ vanish, leaving a pure $T\varphi^3$ term in the potential. The barrier can be increased by adding more scalar degrees of freedom and by choosing their mass parameters to roughly cancel the thermal contributions: $m_k^2\sim -\Pi_k(T)$. Since the physical masses are given by $m^2\sim m_k^2+y_k v^2$, a SFOEWPT can lead to relatively light degrees of freedom at $T=0$ under this scenario. The most widely studied example occurs in the MSSM (see, {\em e.g.} Ref.~\cite{Carena:1997ki}), where the right-handed stop mass parameter is chosen to cancel its thermal mass, leading to a stop that is lighter than the top quark. The effect is particularly important for stops, since they have large Yukawa couplings and introduce a factor of $N_C$ when the sum in Eq.~(\ref{eq:deltav}) is performed.

Searches for light stops at the LHC\cite{Aad:2014kra,Chatrchyan:2013xna,CMS-PAS-SUS-13-009,Khachatryan:2015wza,Aad:2014nra} as well as determinations of Higgs boson signal strengths\cite{Curtin:2012aa,Katz:2015uja} appear to have closed most of the window for a SFOEWPT in the MSSM as well as extensions with hard SUSY-breaking terms. Going beyond SUSY, it was realized that loop contributions involving multiple species may also lower $T_C$ in the presence of a barrier between the two phases\cite{Huang:2012wn}. When one of the $y_k<0$, the corresponding contribution to $\Delta V_\mathrm{daisy}$ will be positive, effectively increasing the energy of the broken phase relative to the unbroken phase. As a result, the Universe must cool to a lower temperature than it otherwise would for the two phases to be degenerate in energy, thereby leading to a lower $T_C$ and increasing the left hand side of Eq.~(\ref{eq:deltav})\footnote{Application of this idea to the stop and sbottom contributions to the effective potential and the implications for the EWPT are under investigation.}.

Possible generation of a SFOEWPT through loop contributions has recently been studied in a general way by the authors of Ref.~\cite{Katz:2014bha}. The introduction of colored scalar particles that lead to a SFOEWPT would also lead to significant increases in the $hgg$ and $h\gamma\gamma$ couplings. Color singlets that are charged under the electroweak gauge groups and that give rise to a SFOEWPT would not modify the $hgg$ coupling but could induce observable deviations in the rate for $h\to\gamma\gamma$. For gauge singlets that generate a SFOEWPT solely via loops rather than tree-level interactions (see below), one would expect a change in $\sigma(e^+e^-\to Zh)$. The projected sensitivity of the CEPC and FCC-ee could allow for an observation of these modifications of Higgs boson production and decays\footnote{We also note that loop effects arising from new degrees of freedom may modify the $T=0$ potential in such a way that the SM gauge boson finite-temperature loops induce a SFOEWPT.}.


\subsection{Tree-level interactions}
\label{sec:tree}
The introduction of additional Higgs boson-scalar interactions leads to a number of possibilities for a SFOEWPT, including new patterns of EWSB where the occurrence of multi-step transitions to the present electroweak phase may entail such transitions. At the level of renormalizable operators, a broad range of possibilities are embodied in the Higgs portal interactions of Eq.~(\ref{eq:portal1}). 

The simplest scenario arises when $\phi$ is a real singlet, denoted here as $S$. At $T=0$, the presence of the two operators in Eq.~(\ref{eq:portal1}) implies the existence of two mass eigenstates $h_{1,2}$ that are doublet-singlet mixtures\footnote{We will take $h_1$ to be the SM-like Higgs scalar.}. For the dynamics of the transition, taking $a_1<0$, the cubic interaction introduces a tree-level barrier, generating a non-vanishing $e>0$. The presence of a non-vanishing $a_2$ may lead to a reduction in the value of ${\bar\lambda}$ and a reduction in $T_C$. 
For $a_2<0$, one finds a direct reduction in ${\bar\lambda}$\cite{Profumo:2007wc}. For positive $a_2$, the effect is indirect, involving the interplay of parameters in the scalar mass-squarked matrix\cite{Profumo:2014opa}. These features have been analyzed in a general fashion in Refs.~\cite{Profumo:2007wc,Espinosa:2011ax,Profumo:2014opa,Damgaard:2013kva,Curtin:2014jma}, while specific model realizations in the NMSSM have been studied in Refs.~\cite{Menon:2004wv,Kozaczuk:2014kva}.

\begin{figure}[ht!]
  \centering
  \includegraphics[width=0.5\linewidth]{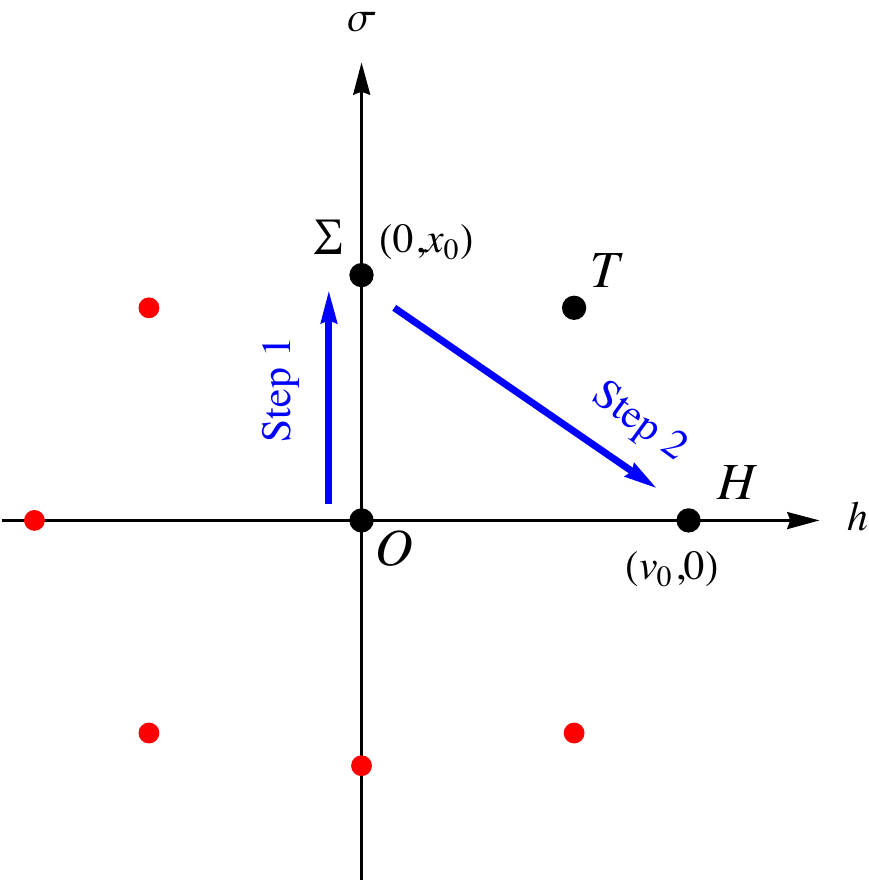}
  \caption{\label{fig:sigmasm} Two-step electroweak symmetry breaking for the real triplet Higgs portal scenario~\cite{Patel:2012pi}. Here, $h$ and $\sigma$ denote the neutral Higgs boson and real triplet background fields, respectively; $O$ gives the location of the symmetric phase, while $\Sigma$ and $H$ denote the locations of the triplet and Higgs phases, respectively. $T$ indicates another extremal point, while the red points are related to the black points by $Z_2$ symmetries. The two step transition proceeds first along the triplet direction at a temperature $T_\sigma$, followed by a transition to the Higgs phase at a temperature $T_h< T_\sigma$. }
\end{figure}

For $\phi$ transforming non-trivially under SU(2$)_L$, the constraints on the electroweak $\rho$-parameter imply that the vev of the neutral component of $\phi$ must be small at $T=0$. As the latter is proportional to $a_1$, the corresponding tree-level barrier induced by non-vanishing $a_1$ between the symmetric  and present electroweak vacua is too small to allow for a SFOEWPT at finite-$T$. On the other hand, for suitable choices of parameters, it is possible that electroweak symmetry breaks in two steps: (1) at temperature $T_1$, the neutral component of $\phi$ gets a vev while the doublet vev vanishes; (2)  at $T_2< T_1$, a second transition occurs to the vacuum with vanishingly small neutral $\phi$ vev and non-vanishing doublet vev. The first transition may be strongly first order, and in this phase sphaleron transitions are suppressed since $\phi$ transforms non-trivially under SU(2$)_L$. The matter-antimatter asymmetry may be produced during this first step and preserved during the transition at $T_2$. A concrete realization of this possibility for $\phi$ being a real SU(2$)_L$ triplet has been analyzed in Ref.~\cite{Patel:2012pi,Inoue:2015pza}, while general considerations have been discussed in Ref.~\cite{Blinov:2015sna}. The vacuum structure of the potential and the two-step trajectory is illustrated in Fig.~\ref{fig:sigmasm}.

For the special case of the two Higgs doublet model (2HDM) where the $\rho$-parameter constraints do not apply to the vevs, initial studies by the authors of Refs. \cite{Dorsch:2013wja,Dorsch:2014qja} indicate that it is possible to achieve a SFOEWPT in a singlet step transition due to the presence of additional tree-level interactions involving the two doublets. The SFOEWPT-favored region of parameter space suggests that the neutral pseudoscalar $A^0$ should be moderately heavy and that it may be considerably heavier than the non-SM-like neutral scalar $H^0$. 


Relaxing the requirement of renormalizability, the existence of higher-dimensional Higgs self-interactions may also enable a SFOEWPT during a singlet-step transition to the present electroweak vacuum\cite{Delaunay:2007wb,Grinstein:2008qi,Grojean:2004xa,Bodeker:2004ws,Henning:2014gca}. Writing
\beqn
\label{eq:higherdim}
V(H) = -\mu^2 H^\dag H + \lambda(H^\dag H)^2 + \eta (H^\dag H)^3+\cdots \ \ \ .
\eeqn
one observes that when the dimensional parameter $\eta>0$, the coefficient of the quartic self-coupling may be negative (subject to stability requirements), again introducing a tree-level barrier. This possibility may be particularly interesting when the Higgs is a pseudo-Goldstone boson, so that $\eta\sim 1/F_H^2$, with $F_H$ being  of order 100 GeV \cite{Grinstein:2008qi}. Alternately, when $\phi$ is a gauge singlet that is sufficiently heavy to be integrated out, a potential of the form in Eq.~(\ref{eq:higherdim}) may arise with $\lambda<0$ and $\eta>0$. For suitable choices of the singlet mass and couplings $a_{1,2}$, a SFOEWPT can be achieved. For related work on the EWPT treating the Higgs boson as a pseudo Goldstone Boson, see Ref.~\cite{Grinstein:2008qi}.

\subsection{Higgs portal dark matter}
\label{sec:hpdm}

The literature on dark matter is vast, and this short overview  cannot do justice to the breadth of theoretical work that has been carried out on this subject. From a practical standpoint, we instead focus on a few representative cases that illustrate the features of Higgs portal interactions and that illustrate the phenomenological implications. We consider both scalar dark matter that may arise as part of an extended scalar sector and fermionic dark matter.

\vskip 0.1in
\noindent{\bf Scalar dark matter: gauge singlets}. Extending the SM scalar sector with a real singlet $S$ has been discussed above in the context of the EWPT. To achieve a viable dark matter candidate, the corresponding scalar potential must admit a $Z_2$ symmetry ($S\to -S$) that ensures stability of the $S$, with vanishing $S$ vev at $T=0$.  The universe may still undergo a SFOEWPT to the EWSB Higgs vacuum if the singlet potential contains a tachyonic mass-squared term. In this case, the universe may first land in a vacuum with $\langle H^0\rangle  =0$ and $\langle S \rangle  \not=0$ with decreasing $T$, followed by a transition to the $\langle H^0\rangle  \not=0$ and $\langle S \rangle =0$ vacuum at lower temperature \cite{Curtin:2014jma}. For a non-tachyonic mass-squared term, a SFOEWPT is generally possible only when $\langle S \rangle  \not=0$ at $T=0$. Avoidance of cosmic domain walls then implies that the potential must not be $Z_2$-symmetric, implying that $a_1\not=0$. 

The general case (non-tachyonic mass) has been investigated by a number of authors\cite{Barger:2007im,Cline:2013gha,He:2013suk,Kahlhoefer:2015jma}. In the standard thermal DM paradigm, the relic density is governed by the Higgs portal coupling $a_2$ that sets the strength of the annihilation channels $SS\to hh$ and $SS\to h \to f{\bar f}$ and $VV$. For $m_S$ near $m_h/2$, the annihilation cross section $\sigma_\mathrm{ann}$ is dominated by single Higgs boson exchange (the \lq\lq Higgs pole"), which is resonantly enhanced. The magnitude of $a_2$ must be accordingly reduced in order for the singlet density to saturate the observed DM relic density: $\Omega_S=\Omega_{DM}$. However, if the singlet constitutes one component of a multicomponent dark matter scenario, then one may have a larger magnitude for $a_2$ with $\Omega_S/\Omega_{DM} <1$. Alternately, a non-thermal mechanism may lead to saturation of the observed relic density for larger $a_2$\cite{Feng:2014vea}. Note also that for the tachyonic mass scenario, stability of the Higgs/DM vacuum constrains the DM mass and relic density. The former is given by $-|\mu_s^2| + a_2 v^2$. The coupling $a_2$ must be positive and sufficiently large as to ensure a stable Higgs/DM vacuum. However, increasing $a_2$ leads to a smaller $\sigma_\mathrm{ann}$. For a detailed study, see, {\em e.g.} Ref.~\cite{Gonderinger:2009jp}.  

DM direct detection experiments constrain the product $(\Omega_S/\Omega_{DM}) \times \langle \sigma_\mathrm{ann} v \rangle$, $v$ is the DM velocity. Constraints obtained including the recent LUX results are shown in Fig. 9 of Ref.~\cite{Curtin:2014jma} , assuming a thermal DM scenario. For non-thermal DM that yields $\Omega_S/\Omega_{DM} =1$ the constraints are considerably more severe, except for $m_S$ near $m_h/2$ . The corresponding collider signatures are generally quite challenging (for a recent discussion, see, {\em e.g.}, Ref.~\cite{Craig:2014lda}). Since $\langle S \rangle =0$, the singlet cannot be produced directly in $pp$ collisions as it does not mix with the SM Higgs boson and does not couple to any SM fields except pairwise to the Higgs boson. For $m_S \leq m_h/2$, the Higgs portal interaction $a_2$ leads to an invisible decay mode for the Higgs boson. The present LHC upper bound on the invisible branching fraction BR$(h\to\mathrm{invis})$ is 23\%~\cite{Aad:2015pla,Khachatryan:2014jba}. Note that this bound constrains the singlet fraction of the DM relic density from being arbitrarily small, since $|a_2|$ cannot be arbitrarily large. For $m_S > m_h/2$, $S$ is pair produced only through off-shell Higgs boson processes that have a relatively small cross section. 

%
 %
%
%
%
%

Variants on the simplest scenario of the SM plus a real singlet include extending the SM with a complex singlet\cite{Barger:2008jx,Gonderinger:2012rd,Jiang:2015cwa} and extending the 2HDM with a real singlet\cite{He:2013suk,Drozd:2014yla}. As discussed in Ref.~\cite{Barger:2008jx} for the complex singlet $\mathbb{S}$, it is convenient to assign to the singlet field a global U(1) charge. For a U(1) conserving potential, $\mathbb{S}$ is equivalent to two degenerate real singlets that  behave much like the real singlet as a dark matter candidate. Introducing U(1)-breaking into the potential then splits the masses of the two components of $\mathbb{S}$, yielding a two-component DM scenario. The relative contributions of each to the relic density depend on the mass splitting and the magnitude of the singlet quartic self coupling. If $\langle \mathbb{S}\rangle\not=0$, then the Goldstone boson of the spontaneously broken U(1) will be stable. Explicitly breaking the U(1) makes this field a massive, viable DM candidate. The remaining degree of freedom associated with $\mathbb{S}$ then behaves like the real singlet that mixes with the SM Higgs boson. The resulting two mass eigenstates then appear as s-channel poles in the DM annihilation amplitude. The DM direct and indirect detection signatures, as well as the collider phenomenology, entails a combination of those associated with the real singlet when it is either a DM candidate or unstable. Under this scenario, one may also encounter a SFOEWPT.
As also noted in Ref.~\cite{Barger:2008jx}, the presence of the complex scalar then allows for complex couplings in the potential as well as a situation in which both components of the scalar receive vevs, a feature that would generally preclude the existence of a viable DM candidate. However, if that situation is associated with the first step of a two-step EWPT, and if it goes to zero during the second transition to the SM vacuum, then one (or more) of the components of  $\mathbb{S}$ may remain stable at and below the freeze-out temperature for thermal DM\cite{Jiang:2015cwa}.

Another variant on the simplest scenario of the SM plus a real singlet has been studied recently in Refs.~\cite{He:2013suk,Drozd:2014yla} using a 2HDM plus a real singlet. Generically, the presence of a second neutral CP-even Higgs state opens up the parameter space of Higgs portal couplings that is consistent with the observed relic density. For low values of $m_S$, the annihilation rate to $b{\bar b}$ can be enhanced for large $\tan\beta$ in the Type II 2HDM, further opening the available parameter space, thereby allowing for a value of $a_2$ consistent with direct detection constraints and while achieving the observed relic density\cite{Drozd:2014yla}.

\vskip 0.1in 
\noindent{\bf Non-singlet scalar dark matter: minimal dark matter}. When $\phi$ carries SM gauge charges and has integer isospin, the neutral component $\phi^0$ may be a viable DM candidate. A similar situation holds for one of the neutral states in a 2HDM. Stability of $\phi^0$ requires a $\mathbb{Z}_2$ symmetry, which must either be imposed by hand for lower dimensional representations of SU(2$)_L$ or is automatic for higher dimensional representations. The latter situation generally corresponds to \lq\lq minimal dark matter" \cite{Cirelli:2005uq,Cirelli:2007xd}. In order to evade DM direct detection limits, scalar dark matter must either have zero hypercharge or a highly suppressed coupling to the $Z^0$ boson; otherwise, the DM-nucleus neutral current scattering cross section will lie well above present direct search bounds. 

A general study of non-singlet scalar DM has been carried out in Ref.~\cite{Hambye:2009pw}. These authors found that the both the $\mathbb{Z}_2$ symmetric 2HDM and both the real or complex scalar multiplets of dimension $n=3$, 5, or 7 may be viable. The lower bounds on the corresponding masses range from roughly 500 GeV to 20 TeV depending on $n$ and assuming saturation of the relic density. A detailed study of the collider phenomenology for the real $n=3$ case has been carried out in Ref.~\cite{FileviezPerez:2008bj}. In general one would search for one or more disappearing charged particle tracks when one or more charged states is produced through electroweak pair production. Results of a CMS search for disappearing charged particle tracks are reported in Ref.~\cite{CMS:2014gxa}. For neutralinos in the MSSM under the anomaly mediated SUSY-breaking scenario, a neutralino with mass less than 260 GeV is excluded. One might anticipate a similar bound for the scalar DM scenarios, though to our knowledge none has yet been reported in the literature. For the real triplet of Ref.~\cite{FileviezPerez:2008bj}, saturation of the full relic density requires a DM mass at or above roughly 2 TeV, well above the present CMS exclusion. Note that in the two-step EWPT scenario described earlier (see Sec.~\ref{sec:tree}), the CMS exclusion would likely preclude a stable neutral triplet in the minimal version of this scenario. 

\vskip 0.1in
\noindent{\bf Fermionic dark matter}. In general, a Higgs portal interaction with one or more additional fermionic multiplets $\chi$ may provide for a viable fermionic DM scenario, though it will not substantially affect the EWPT. On the other hand such an interaction may provide for new sources of CPV as needed for EWBG during a SFOEWPT, a possibility that has been analyzed recently in Refs.~\cite{Chao:2014dpa,Chao:2015uoa} but that we will not explore in detail here. For the simplest, pure DM scenario, one introduces a non-renormalizable interaction of the form
\begin{equation}
\label{eq:fermionic}
\mathcal{L} \supset \frac{1}{\Lambda} H^\dag H\left( \cos \xi\, {\bar\chi} \chi + \sin\xi\, {\bar\chi} i\gamma_5 \chi\right)\
\end{equation}
where $\chi$ is a SM gauge singlet and $\Lambda$ defines an effective mass scale that incorporates the values of the scalar and pseudoscalar operator coefficients up to their relative normalization. A detailed study of this scenario has been carried out in Ref.~\cite{Fedderke:2014wda}. Saturation of the relic density requires $\Lambda$ to range from a few hundred GeV to a few TeV, while constraints on the Higgs boson invisible width impose a strong exclusion for  $m_\chi<m_h/2$. Unitarity constraints on an extension of the simplest scenario that includes both the SM Higgs boson and a real scalar singlet have been studied in Ref.~\cite{Walker:2013hka}. Generation of the baryon asymmetry in a 2HDM variant of the interaction (\ref{eq:fermionic}) has been studied in Ref.~\cite{Chao:2015uoa}. 

\vskip 0.1in
\noindent{\bf Portal to a hidden sector}. Dark matter particles may exist in a hidden sector containing additional degrees of freedom that may couple to the Higgs boson. If these degrees of freedom are sufficiently light, they may significantly affect dark matter dynamics,  leading, for example, to an enhanced annihilation cross section. A particularly interesting possibility is that the hidden sector contains a U(1) symmetry that is spontaneously broken. The corresponding massive gauge boson $X$ may mix with the SM hyper charge boson, leading to the existence of an additional \lq\lq dark photon" or \lq\lq dark Z", $Z_D$. Alternately, the hidden sector may contain one or more SM gauge singlet scalars $S$ that couple to the $X$. As discussed in Section \ref{sec:tree}, Higgs portal interactions of the form (\ref{eq:portal1}) will lead to $h$-$S$ mixing. In this context, the Higgs-like state $h_1$ will inherit the couplings of $S$ to $X$, albeit suppressed by the mixing angle $\sin\theta$. For $Z_D$ mass below $m_h/2$, this scenario leads to the exotic Higgs decay $h_1\to Z_D Z_D$. The corresponding phenomenology for this possibility has been extensively analyzed in Ref.~\cite{Curtin:2014cca}. Rather than repeat that discussion here, we refer the reader to that work. 

\vskip 0.1in
\noindent A summary of various scenarios is given in Table \ref{tab:scenarios}.

\begin{table}[hb]
\small
\begin{centering}
\begin{tabular}{|c|lc|c|c|c}
\hline\hline
Scenario	&  BSM  & EWPT & DM & Collider Signatures \\
& DOF &&&\\
\hline\hline
Real Singlet		 & 1 & Tree-level  & $X$ & $\ast$ Two neutral mixed states\\ 
&& single-step && $\ast$ Modified trilinear self-coupling \\
&&or two-step && $\ast$ Reduced SM-like signal strength\\
&&&& $\ast$ Resonant di-Higgs production\\
&&&& w/ final states: $b{\bar b}\tau^+\tau^-$, $b{\bar b}\gamma\gamma$ {\em etc.}\\
&&&& $\ast$ Exotic decays: $f{\bar f} \gamma \gamma$ {\em etc.}\\
\hline
Real Singlet 		& 1 & $X$ & $\surd$ & $\ast$ Invisible Higgs decays \\
\hline
Real Singlet & 1 & Two-step & $\surd$ & $\ast$ VBF $h^\ast\to$ invisible\\
&&Loop-induced&&$\ast$ Modified trilinear self-coupling \\
&& one-step&&\\
\hline\hline
Complex Singlet	& 2 & Tree-level  & $\surd$&$\ast$ Two neutral mixed states  \\
&& single-step &&plus scalar DM\\
&&  &&$\ast$ Modified trilinear self-coupling\\
&&&& $\ast$ Reduced SM-like signal strength\\
&&&& $\ast$ Resonant di-Higgs production\\
&&&& w/ final states: $b{\bar b}\tau^+\tau^-$, $b{\bar b}\gamma\gamma$ {\em etc.}\\
&&&& $\ast$ Exotic decays: $f{\bar f} \gamma \gamma$ {\em etc.}\\
&&&& $\ast$ Invisible decays\\
\hline\hline
Real Triplet		& 3 & Loop-induced  & $\surd$ & $\ast$ Reduced $\Gamma(h\to\gamma\gamma)$ \\
&& multi-step && $\ast$ Disappearing charged particle tracks \\
\hline
Real Triplet		& 3 & Loop-induced  & $X$ & $\ast$ Reduced $\Gamma(h\to\gamma\gamma)$ \\
&& multi-step && $\ast$ EW $H^\pm H^\mp$ and $H^\pm H^0$ production\\
&&&& w/ novel final states: $b{\bar b}\tau\nu$, $b{\bar b} W^\pm Z$, {\em etc.}\\
\hline\hline
2HDM			& 4 & Loop-induced & $X$ & $A^0\to Z^0 H^0$ \\
&& single step && \\
\hline\hline
Higher-dim  & 0 & Tree-level & $X$ & \\
operators&& single-step &&\\
\hline\hline
\end{tabular}
\caption{Higgs portal scenarios. For each scenario, the number of additional BSM degrees of freedom (DOF) is indicated, along with the nature of the EWPT and viability as a contributor to the dark matter (DM) relic density. The quantity $\surd$ ($X$) indicates viable (not viable). The final column summarizes the possible collider signatures for each scenario.\label{tab:scenarios}}
\end{centering}
\end{table}

\section{Signatures and Benchmarks}
\label{sec:signatures} 
Because the nature of an EWPT depends critically on the interactions of new particles with the Higgs doublet, measuring the Higgs boson properties, including its self-interactions as well as interactions with other SM particles through production and decays can provide a probe of phase transition dynamics. In addition, it is possible to search for the new degrees of freedom through their direct production. Thus,  the scenarios for baryogenesis-favorable phase transitions summarized in Section \ref{sec:theory} may lead to a number of signatures accessible at the LHC and prospective future colliders. For each of the simplest Higgs portal scenarios, the possibilities are indicated in the last column of Table \ref{tab:scenarios}. We also indicate which scenarios may allow for a viable dark matter candidate and indicate some of the corresponding signatures.

In what follows, we summarize work completed to date on some of these prospective signatures, followed by a brief roadmap for future analyses.

\subsection{Modified Higgs self-coupling}
\begin{figure}[ht!]
  \centering
  \includegraphics[width=0.5\linewidth]{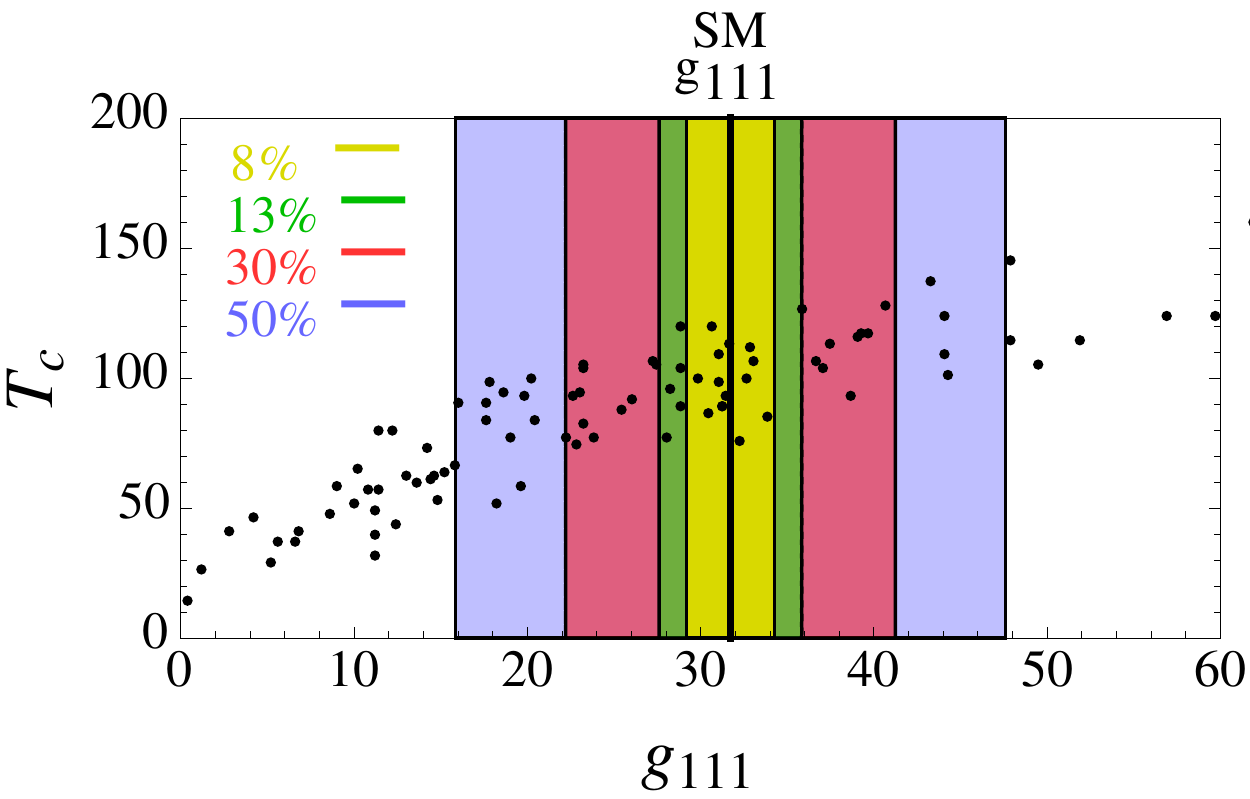}
  \caption{\label{fig:selftc} Correlation between the critical temperature and SM-like Higgs scalar self-coupling in the singlet-extended SM, adapted from Ref.~\cite{Profumo:2014opa}. Colors indicate future sensitivities of the HL-LHC (purple), CEPC/FCC-ee (red), ILC (green), and SPPC/FCC-hh (yellow).}
\end{figure}

In the real singlet Higgs portal scenario, the Higgs cubic self interaction $\lambda v h^3$ becomes $g_{111} h_1^3$, where $g_{111}$ is a combination of the potential parameters, $v$, and the singlet vev. The value of this coupling can provide a probe of a SFOEWPT\cite{Noble:2007kk,Katz:2014bha,Profumo:2014opa}. As shown in Fig.~\ref{fig:selftc},  the value of $g_{111}$ is strongly correlated with $T_C$, with smaller $g_{111}$ corresponding to lower critical temperature\cite{Profumo:2014opa}.  Moreover, its value may be considerably less than its pure SM value $\sim 30$ GeV, thereby allowing for rather low transition temperatures. A measurement of $g_{111}$ can, thus, provide a probe of $T_C$ and the SFOEWPT-viable regions of singlet extensions. Speaking conservatively, one expects a $\sim 30-50\%$ determination of this parameter at the HL-LHC\cite{CMS-PAS-FTR-15-002,ATL-PHYS-PUB-2014-019,ATL-PHYS-PUB-2015-046}. Current projections for the ILC anticipate a 27\% and 16\% (10\%) determination using the full data set at $\sqrt{s}=500$ GeV and 2(5) $ab^{-1}$ at 1 TeV respectively, considering the $4b$ and $bbWW$ final states\cite{Fujii:2015jha}. Expectations next generation $pp$ colliders vary. Refs.~\cite{Barr:2014sga,Azatov:2015oxa} project, respectively a 40\% and 30\% determination with 4 $ab^{-1}$ and 3 $ab^{-1}$ at $\sqrt{s}=100$ TeV, implying on the order of a 10\% determination with 30 $ab^{-1}$.
For center of mass energies below the di-Higgs production threshold, as would be relevant for the future circular $e^+e^-$ colliders, one may determine $g_{111}$ indirectly as it contributes to associated production at one-loop order\cite{McCullough:2013rea,Englert:2013tya}. A 30\% determination may be possible with this approach.

\subsection{Modified Higgs boson couplings to SM particles}
\begin{figure}[ht!]
  \centering
  \includegraphics[width=0.5\linewidth]{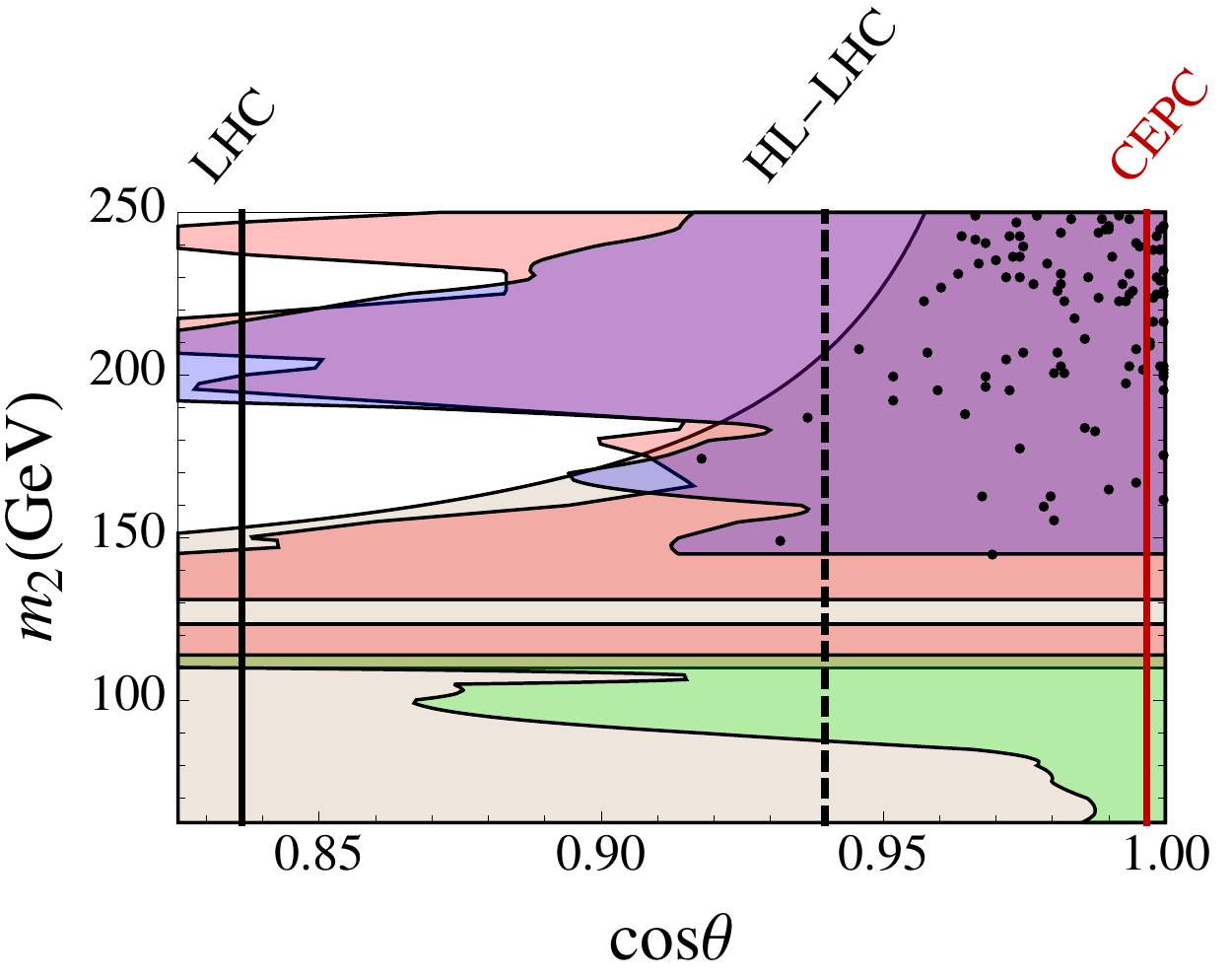}
  \caption{\label{fig:mixewpt} Correlation between the singlet-like Higgs boson mass $m_2$ and doublet-singlet mixing angle $\theta$ in the singlet-extended SM for $m_1=125$ GeV, adapted from Ref.~\cite{Profumo:2014opa}. Black points indicate SFOWEPT-viable models that also satisfy electroweak precision constraints. Colors regions indicate allowed regions from various Higgs boson measurements, heavy Higgs searches, and electroweak precision constraints as discussed in Ref.~\cite{Profumo:2014opa}. Vertical lines indicate mixing angle sensitivities of the present LHC combined fit results and prospective future HL-LHC and CEPC sensitivities. The latter have been obtained assuming a similar sensitivity as would be obtained at the FCC-ee (TLEP 240) collider.}
\end{figure}

The aforementioned scenarios may lead to changes in the Higgs boson couplings to other particles through the effect of Higgs mixing and/or new loop contributions. In the case of doublet-singlet mixing, for example, the SM-like state $h_1$ and singlet-like state $h_2$ may be written as
\beqn
\nonumber
h_1& =&h \cos\theta  + S \sin\theta \\
h_2 & = &h \sin\theta  -S \cos\theta \ \ \ ,
\label{eq:mix}
\eeqn
where $|\cos\theta|\geq 1/\sqrt{2}$.
In the regime $m_2> m_1/2$, the SM-like Higgs boson has no new decays and its branching ratios are unchanged from the SM. However, the production cross section, and thus, signal strength, will be reduced by $\cos^2\theta$. Present LHC data imply $\cos^2\theta\gsim 0.71$\cite{Profumo:2014opa}, a bound expected to increase to $\sim 0.95$ with the HL-LHC. Figure~\ref{fig:mixewpt} shows the distribution of parameter space points for a SFOEWPT transition in the $\cos\theta$-$m_2$ plane for $2m_1>m_2> m_1/2$. One sees that for EWPT-viable regions of parameter space, there exist considerable possibilities for future precision Higgs boson studies to observe deviations from signal strengths expected for a purely SM Higgs boson. Some of this parameter space may be probed with the HL-LHC.
A circular $e^+e^-$ Higgs factory, such as the CEPC or FCC-ee,  could allow a probe at better than one percent, and we see that there exist many SFOEWPT-viable models that would correspond to mixing of this magnitude or smaller. 

\begin{figure}[ht!]
  \centering
  \includegraphics[width=0.5\linewidth]{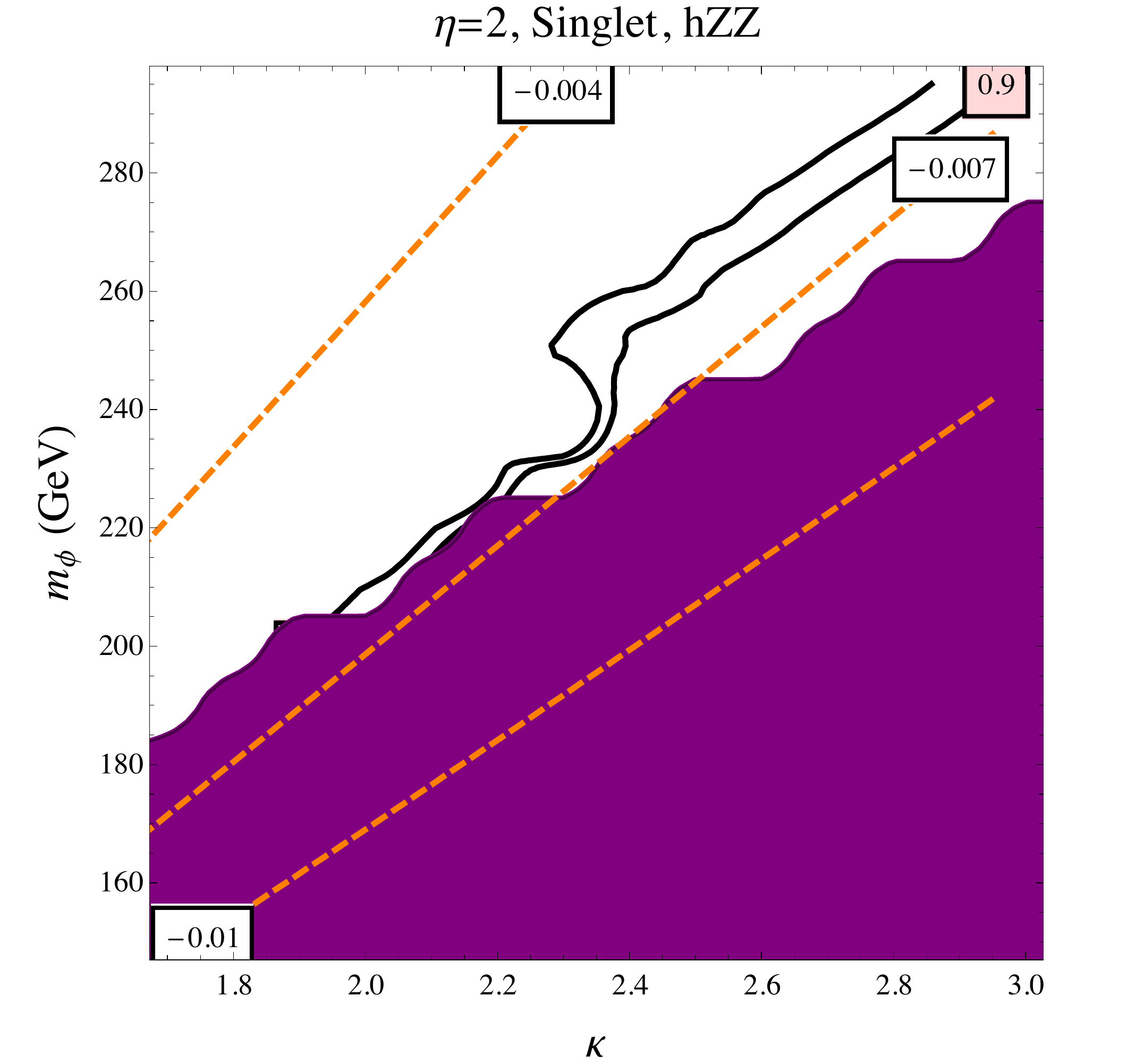}
  \caption{\label{fig:assoc} Correlation between SFOEWPT and $\sigma(e^+e^-\to Zh)$ for the singlet extension of the Standard Model\cite{Katz:2014bha} in the singlet mass-Higgs portal coupling plane. Dashed red lines correspond to relative changes in $\sigma(e^+e^-\to Zh)$. Solid lines give values of the LHS of Eq.~(\ref{eq:sfoewpt}). Purple gives region of non-vanishing singlet vev.}
\end{figure}

For the case where the singlet generates the SFOEWPT solely via loop effects, one may expect an observable reduction in the associated production cross section $\sigma(e^+e^-\to Zh)$ of more than 0.4\% . This possibility is illustrated in Fig.~\ref{fig:assoc}, where the dashed lines indicate the relative change in the associated production cross section and the solid lines indicate the value of the RHS of Eq.~(\ref{eq:sfoewpt}).  In the regions where the RHS is $\gsim 1$, the reduction in $\sigma(e^+e^-\to Zh)$ is expected to be large enough to be observed at the CEPC or FCC-ee. 

\begin{figure}[ht!]
  \centering
  \includegraphics[width=0.8\linewidth]{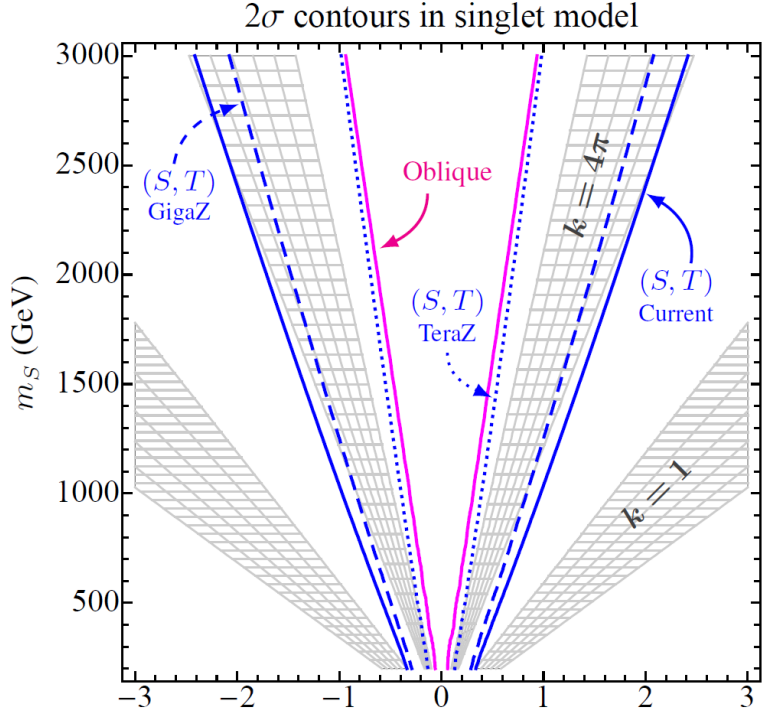}
  \caption{
  \(2\sigma\) Contours of future precision measurements for a heavy singlet-catalyzed EWPT\cite{Henning:2014gca} in the $(a_1/m_S, m_S)$ plane. The magenta contour is the \(2\sigma\) sensitivity to the universal Higgs oblique correction entering associated production at the ILC 500up. Blue contours show the \(2\sigma\) RG-induced constraints from the \(S\) and \(T\) parameters  from current measurements (solid) and future sensitivities at ILC GigaZ (dashed) and TLEP TeraZ (dotted). Regions of a viable first order EWPT, from Eq.~(\ref{eq:heavyewpt}), are shown in the gray, hatched regions for \(k\equiv a_2 =1\) and \(4\pi\). Figure courtesy of the authors of Ref.~\cite{Henning:2014gca}. \label{fig:heavysinglet}}
\end{figure}

When the singlet mass $m_S$ parameter is heavy ($m_S >> m_h$), one may integrate out the singlet degrees of freedom, thereby generating the higher dimension operators\cite{Henning:2014gca,Gorbahn:2015gxa} 
\beqn
\mathcal{O}_H & = & \frac{1}{2}\left(\partial_\mu H^\dag H\right)^2 \\
\mathcal{O}_6 & = &  (H^\dag H)^3\ \ \ .
\eeqn
The resulting contribution to the Lagrangian will be
\begin{equation}
\mathcal{L}_\mathrm{eff}\supset \frac{a_1^2}{m_S^4}\mathcal{O}_H-\left( \frac{a_1^2 a_2}{m_S^4}-\frac{2 a_2^3 b_3}{m_S^6}\right)\mathcal{O}_6
\end{equation}
where $b_3/3$ is the singlet cubic self coupling. The resulting universal Higgs coupling correction  is
\begin{equation}
\delta Z_h = \frac{2 v^2 a_1^2}{m_S^4}
\end{equation}
independent of the quartic Higgs portal coupling $a_2$. On the other hand, both the cubic and quartic portal couplings can enable SFOEWPT if
\begin{equation}
\label{eq:heavyewpt}
\frac{2 v^4}{m_H^2} < \frac{m_S^2}{a_1^2 a_2} < \frac{6 v^4}{m_H^2}\ \ \ .
\end{equation} 
The resulting constraints on the $(a_1, m_S)$ parameter space for different representative choices for $a_2$ are shown in Fig.~\ref{fig:heavysinglet}. The gray regions correspond to a SFOEWPT for different representative values of $a_2$. The region below the pink solid lines could be probed with the measurement of $\sigma(e^+e^-\to Zh)$ at the CEPC/FCC-ee. Additional constraints arising from present and possible future electroweak precision tests are induced by the blue lines. 

\begin{figure}[ht!]
  \centering
  \includegraphics[width=0.5\linewidth]{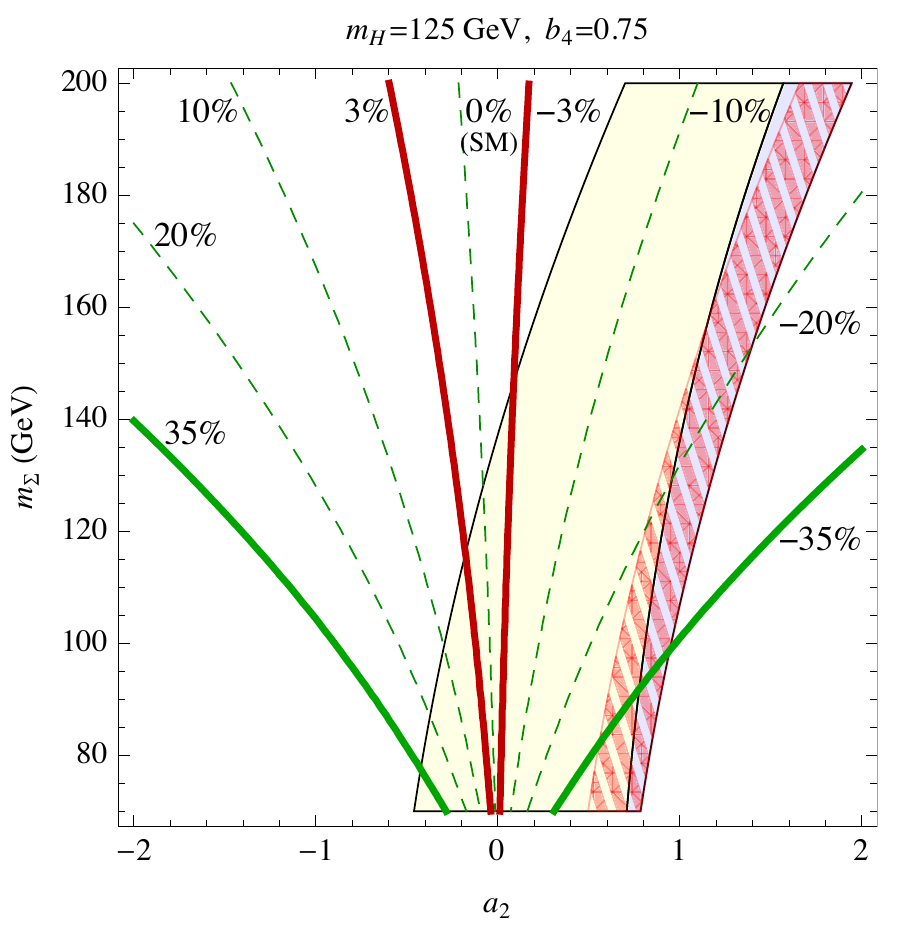}
  \caption{\label{fig:triplethgg} Shift in $\Gamma(h\to\gamma\gamma)$ in the real triplet extended SM~\cite{Patel:2012pi}. The red hatched region corresponds to a two-step phase transition. Dashed curves indicate relative changes in $\Gamma(h\to\gamma\gamma)$ compared to its Standard Model value. Solid green and red curves correspond to relative shifts of $\pm 35\%$ and $\pm 3\%$, respectively. Figure courtesy of H. Patel.}
\end{figure}

For the Higgs portal scenarios involving SU(2$)_L$ non-singlet representations $\phi$, loop contributions may also lead to modified Higgs boson couplings to SM particles. If $\phi$ is an SU(3$)_C$ singlet, one will expect no modification of the gluon fusion operator $H^\dag H GG$, but new contributions to the rate for $h\to\gamma\gamma$ will appear whose impact will depend on the magnitude and sign of $a_2$ and the $\phi$ mass. For the real triplet illustration, the correlation between the relative change in $\Gamma(h\to\gamma\gamma)$, the values of $a_2$ and the triplet mass, and the occurrence of a two-step transition are shown in Fig.~\ref{fig:triplethgg}. One finds that the existence of the two-step transition generally implies a reduction in the Higgs diphoton decay rate. With the HL-LHC, the CMS collaboration projects a probe of the diphoton signal strength with 2\% precision\cite{CMS:2013xfa}, while ATLAS projects at least a 5\% determination\cite{ATL-PHYS-PUB-2014-016}. Combining the projected LHC determination of the ratio of branching ratios $\mathrm{BR}(h\to\gamma\gamma)/\mathrm{BR}(h\to ZZ)$ with the precise determination of the $ZZh$ coupling at the ILC would allow a one percent determination of the Higgs diphoton coupling\cite{Fujii:2015jha}. The present CEPC projection is for a 4.7\% determination of the coupling with 5 ab$^{-1}$ integrated luminosity\cite{CEPCpCDR}, while for the FCC-ee one anticipates  3\% determination of the signal strength with associated production and 10 ab$^{-1}$ of data\cite{Gomez-Ceballos:2013zzn}. Determinations at this level of precision would provide a decisive test of the simplest realization of the multi-step transition paradigm.


\subsection{Exotic Higgs boson Decays}

The discovery of non-standard (exotic) Higgs boson decays as well as of non-SM Higgses could provide important information about Higgs portal interactions.  As indicated above, the decay $h\to Z_D Z_D$ would signal the presence of a new dark sector containing a SM gauge singlet scalar and a U(1)$^\prime$ gauge boson\cite{Curtin:2014cca}. For $m_S < m_h/2$, the SM Higgs boson may decay invisibly to a pair of $S$ bosons.  In the absence of $h$-$S$ mixing, this decay mode will correspond to an invisibly decaying Higgs boson, whereas in the presence of $h$-$S$ mixing the final states will involve pairs of SM Higgs boson decay final states.

For the non-SM neutral Higgses, which appear in the new physics models with an extended Higgs sector, most of the current searches at the LHC focus on the conventional Higgs boson search channels of $WW$, $ZZ$, $\gamma\gamma$, $\tau\tau$ and $bb$ channel  \cite{Aad:2014vgg,Khachatryan:2015qxa,CMS:2014cdp,Chatrchyan:2013qga,ATLAS-CONF-2013-027,Khachatryan:2014jya,Aad:2014kga,Khachatryan:2014wca,ATLAS-CONF-2013-090,CMS-PAS-HIG-13-021}.   The production of the extra Higgses is usually suppressed compared to that of the SM Higgs boson, either due to its larger mass or its suppressed couplings to the SM particles.  The decay of the heavy neutral Higgses to the $WW$ and $ZZ$ final states is absent for the CP-odd Higgs boson, and could be highly suppressed for the non-SM like CP-even Higgs boson.    The decay modes of $\tau\tau$ or $bb$    suffer from either suppressed signal or large SM backgrounds, and are therefore only relevant for regions of the parameter space with an enhanced $bb$ or $\tau\tau$ coupling.    The search for the charged Higgs bosons at the LHC is even more difficult.  For $m_{H^\pm}>m_t$, the cross section for the dominant production channel of $tbH^\pm$ is typically small.  The dominant decay mode $H^\pm \rightarrow tb$  is hard to identify given the large $tt$ and $ttbb$ background,  while the subdominant decay of $H^\pm \rightarrow \tau\nu$ has suppressed branching fraction.    In the MSSM, even at the end of the LHC running, there is a ``wedge region''~\cite{Dawson:2013bba} in the $m_A-\tan\beta$ plane for $\tan\beta\sim 7$ and $m_A\gtrsim 300$ GeV in which only the SM-like Higgs boson can be covered at the LHC.   Similarly, the reach for the non-SM Higgses is limited in models with an extended Higgs sector.

 \begin{table}
 {\small 
 \begin{tabular}{l|l|l|l} \hline
 &Decay&Final States&Channels  \\ \hline  
 &$HH$ type & $(bb/\tau\tau/WW/ZZ/\gamma\gamma)(bb/\tau\tau/WW/ZZ/\gamma\gamma)$ & $\H \rightarrow \A \A$, ... \\   \cline{2-4}
Neutral Higgs&$HZ$ type & $(\ell\ell/qq/\nu\nu)(bb/\tau\tau/WW/ZZ/\gamma\gamma)$ & $\H \rightarrow \A Z, \A \rightarrow  \H Z$, ...  \\   \cline{2-4}
$\H$, $\A$& $H^+H^-$ type & $(tb/\tau\nu/cs)(tb/\tau\nu/cs)$ & $\H  \rightarrow H^+H^-$, ...  \\  \cline{2-4}
&$H^\pm W^\mp$ type & $(\ell\nu/qq^\prime)(tb/\tau\nu/cs)$ & $\H/\A \rightarrow H^\pm W^\mp$, ...  \\  \hline
Charged Higgs&$ H W^\pm$ type & $(\ell\nu/qq^\prime)(bb/\tau\tau/WW/ZZ/\gamma\gamma)$ & $\hc \rightarrow  \H W, \A W$, ...  \\  \hline
 \end{tabular}
}
 \caption{Exotic decay modes for  Higgses in the 2HDM with the heavy Higgs boson decays to two light Higgses or one light Higgs boson with one SM gauge boson.   $H$ in column two refers to any of the neutral Higgs bosons $\h$, $\H$ or $\A$.  }
 \label{tab:decay_exo}
 \end{table}

In addition to their decays to the SM particles, non-SM Higgses can decay via exotic modes, {\em i.e.}, heavier Higgs decays into two light Higgses, or one light Higgs boson plus one SM gauge boson.   Five main exotic decay categories for Higgses of the Type II 2HDM are shown in Table~\ref{tab:decay_exo}.   These channels typically dominate once they are kinematically open.   A recent study on   exotic Higgs boson decays can be found in  Refs.~\cite{Coleppa:2013xfa,Coleppa:2014hxa,Brownson:2013lka, Coleppa:2014cca, AW_light,Hpm_Tong,Maitra:2014qea,Basso:2012st,Dermisek:2013cxa,Mohn:833753,Assamagan:2000ud,Assamagan:2002ne}.


Note that most of the current experimental searches for the non-SM Higgs boson  assume the absence of exotic decay modes.  Once there are light Higgs states such that these exotic modes are kinematically open, the current search bounds can be  greatly relaxed given the suppressed decay branching fractions into SM final states~\cite{Coleppa:2014hxa,Coleppa:2014cca,Hpm_Tong}.  Furthermore, exotic Higgs decays to final states with two light Higgses or one Higgs boson plus one SM gauge boson provide complementary search channels.  Here, we list such exotic Higgs decays and consider potential search strategies.

\begin{itemize}
\item{$\H\rightarrow \A \A$ or $\H\rightarrow \h \h$}
\end{itemize}
With the final state Higgs boson decaying via $bb$, $\gamma\gamma$, $\tau\tau$, $WW^*$, final states of $bbbb$, $bb\tau\tau$, $bb\gamma\gamma$ and $\gamma\gamma WW^*$ can be used to search for resonant Higgs boson decay to two light neutral Higgses.    Current searches at the LHC 8 TeV with about 20 ${\rm fb}^{-1}$ luminosity  observed 95\% C.L. cross section limits of   2.1 pb at 260 GeV and about 0.018 pb at 1000 GeV~\cite{Aad:2015xja} (see also Refs.~\cite{Aad:2015uka,Aad:2014yja,Khachatryan:2015yea}).  While $bb\gamma\gamma$ and $bb\tau\tau$ have comparable sensitivities at low mass, the $bbbb$ mode dominates at high mass.

\begin{itemize}
\item{$\H\rightarrow \A Z$ or $A \rightarrow \H Z $}
\end{itemize}
With $Z \rightarrow \ell\ell$ and $\H /\A \rightarrow bb, \tau\tau$, the final states of $bb\ell\ell$, $\tau\tau\ell\ell$ can be obtained with gluon fusion production, or in the $bb$ associated production with two additional $b$ jets~\cite{Coleppa:2013xfa,Coleppa:2014hxa,Brownson:2013lka}.   Recent searches from ATLAS and CMS have shown certain sensitivity in this channel~\cite{Khachatryan:2015lba,CMS-PAS-HIG-14-011,Khachatryan:2014jya,Aad:2015wra,CMS-PAS-HIG-15-001}.   In parameter regions where ${\rm Br}(A \rightarrow \H Z)\times {\rm Br}(\H \rightarrow Z Z)$ is not   completely suppressed, $ZZZ$ final states with two $Z$ decaying leptonically and one $Z$ decaying hadronically can also be useful~\cite{Coleppa:2014hxa}.   Other channels with top final states could  be explored as well.

Note that the decay $A\to Z \H$ has been identified as a particular signature of a SFOEWPT in the 2HDM\cite{Dorsch:2013wja}. As discussed below, the prospects for observing this channel in the $\ell\ell b{\bar b} $ and $\ell\ell W^+ W^-$ model have been analyzed in Ref.~\cite{Dorsch:2014qja}.

\begin{itemize}
\item{$\H \rightarrow H^+ H^- $}
\end{itemize}
With both $H^\pm$ decaying via $\tau\nu$ final states, the signal of $\tau\tau \nu \nu$ can be separated from the SM $W^+W^-$ background since the charged tau decay product   in the signal typically has a hard spectrum compared to that of the background~\cite{Hpm_Tong}.      

\begin{itemize}
\item{$\H/\A \rightarrow  H^\pm W^\mp$ }
\end{itemize}

Similar to the $H^+H^-$ case, $H^\pm \rightarrow \tau \nu, tb$ and $W\rightarrow \ell \nu$ with $\ell \tau \nu \bar\nu$  or $tb\ell \nu$ could be used to search for $\H/\A \rightarrow  H^\pm W^\mp$.   Note that for the CP-even Higgs boson $\H$, the branching fraction of $\H \rightarrow H^\pm W^\mp$ is mostly suppressed comparing to $\H \rightarrow H^+H^-$ as long as the latter process is kinematically open and not accidentally suppressed~\cite{Hpm_Tong}.  However, for the CP-odd Higgs boson $\A$, this is the only decay channel with a charged Higgs boson in the decay products.

\begin{itemize}
\item{$\hc \rightarrow \H W^\pm, \A W^\pm $}
\end{itemize}
This is the only exotic decay channel for the charged Higgs boson in the 2HDM.  Given the associated production of $tbH^\pm$, and the decay of $\H$, $A$ into the $bb$ or $\tau\tau$ channel, $\tau\tau bb WW$ or $bbbbWW$  can be used to probe this channel~\cite{Coleppa:2014cca}.  $\H/\A \rightarrow t\bar{t}$ could also be used given the boosted top in the high energy environment.

Note that while  $H^\pm \rightarrow WZ$ is absent in 2HDM type extension of the SM Higgs sector, it could appear, however,  in the real triplet models extension of the SM once the triplet obtains a vev~\cite{FileviezPerez:2008bj}.

\begin{itemize}
\item{$A/\H\rightarrow tt$, $H^\pm \rightarrow tb$}
\end{itemize}

While $A/\H\rightarrow tt$ and $H^\pm \rightarrow tb$ are considered to be challenging experimentally,  a  recent study of BSM Higgs searches with top quarks in the final states~\cite{Hajer:2015gka} showed that  a combination of the channels of  $pp\to bb \H/A \rightarrow bb tt, bb\tau\tau$, $pp \to \H/A \to tt$ as well as $pp \to tb \hc \rightarrow tbtb, tb \tau\nu$ yields full coverage for $\tan\beta$ and pushes the exclusion limits from the $\mathcal O(1)$ TeV at the LHC to the $\mathcal O(10)$ TeV at a 100 TeV $pp$ collider.

\subsection{New states}
In addition to observing deviations of Higgs boson couplings to SM particles, one may also anticipate direct production of new states associated with SFOEWPT-viable models. Again, the singlet scenario provides the simplest illustration. Its production cross section will be proportional to $\sin^2\theta$, presently constrained to be smaller than $\sim 0.34$. For $m_2< 2 m_1$, the signatures of this state would be identical to those of a SM Higgs boson with reduced signal strength\footnote{Recall that the mixing angle cancels from the branching ratios when no new decay channels are kinematically allowed.}. For $m_2> 2 m_1$, where $m_1$ is the mass of the SM-like Higgs boson, a new decay mode $h_2\to h_1 h_1$ becomes kinematically allowed, introducing the possibility of resonant di-Higgs production. The final states would entail pair-wise combinations of SM Higgs boson decay products, such as $b{\bar b}\gamma\gamma$,
$b{\bar b} \tau^+\tau^-$ {\em etc.}. Recently, the authors of Ref.~\cite{No:2013wsa} showed that observation of resonant di-Higgs production in the 
$b{\bar b} \tau^+\tau^-$ channel -- the same one discussed in Ref.~\cite{Dolan:2012rv} for non-resonant production to determine the self-coupling -- could be feasible with 100 fb$^{-1}$ at the LHC, assuming the present maximal value of 0.34 for $\sin^2\theta$. Other studies have analyzed resonant di-Higgs production with $b{\bar b}\gamma\gamma$ and $4b$ final states, though not with an eye toward the EWPT \cite{Chen:2014ask,Barger:2014taa,Dolan:2012ac}.
Looking to the future, should the constraints on the mixing angle become more severe, higher integrated luminosity, a cleaner background environment, or higher parton luminosity as with the SPPC/FCC-hh would be needed to search for such modes. 

For non-singlet $\phi$, such as the real triplet ($\Sigma$) or 2HDM discussed above, production and decays provide for a rich phenomenology. The electroweak $\rho$-parameter constraints on the non-doublet vevs imply that single scalar production is typically suppressed, and the dominant production mechanism will be electroweak pair production. For multi-Higgs doublet models, on the other hand, neutral scalars may be produced through gluon fusion,  while charged states may be singly-produced through associated production. The final states in regions of interest for cosmology will then depend on the details of each scenario. 

\vskip 0.1in

\noindent {\em Two Higgs doublets}. For the 2HDM, the authors of Ref.~\cite{Dorsch:2014qja} have shown that in the SFOEWPT-viable region of parameter space, the decay $A^0\to Z H^0$ could be discovered at the 14 TeV LHC in the $\ell\ell b{\bar b}$ and $\ell\ell W^+W^-$ channels with integrated luminosity in the $\gsim 20-40$ fb$^{-1}$ range. Analyses of decays involving   charged scalars $H^\pm$ remain to be performed. 

\vskip 0.1in

\noindent {\em Real triplet}. For vanishing triplet vev, the neutral component may contribute to the DM relic density. For masses in the several hundred GeV range, where the two step phase transition appears to be viable, the triplet contribution will comprise less than 10\% of the relic density due to the sizable cross section for annihilation into $W^+W^-$ pairs. At the LHC, triplet states are produced electroweakly in $\Sigma^+\Sigma^-$ and $\Sigma^\pm \Sigma^0$ pairs. Electroweak radiative corrections raise the $\Sigma^\pm$ masses relative to the $\Sigma^0$ mass by $\sim 160$ MeV, so that the $\Sigma\pm$ may decay to a $\Sigma^0$ plus a soft pion or lepton-neutrino pair. None of the SM final state particles will be detectable. The signature would, thus, be large $\met$ plus one or more disappearing charged particle tracks. For a 150 GeV $\Sigma^+$ the $c\tau$ is of order 5 cm\cite{FileviezPerez:2008bj}. The prospective implications of the CMS search for disappearing charged particle tracks\cite{CMS:2014gxa} have been discussed above in Section \ref{sec:hpdm}.

For non-vanishing vev, both the neutral and charged triplet can decay, leading to a rich array of possible final states\cite{FileviezPerez:2008bj}. For $\Sigma^\pm$ masses well below the $W^\pm Z$ threshold, the dominant decay mode is $\tau\nu$ when the triplet vev is of order one GeV, while it is $\Sigma^0$ plus a soft pion for very small vev. The largest $\Sigma^0$ branching fraction is to $b{\bar b}$. Thus, one may anticipate $b{\bar b}\tau\nu$ final states for the first case and $b{\bar b}\gamma\gamma$ or $b{\bar b}\tau^+\tau^-$ for the second.  In the study of Ref.~\cite{No:2013wsa}, it was found that a discovery with the $b{\bar b}\tau\nu$ channel may be possible with 100 fb$^{-1}$ at the 14 TeV LHC, with a significance depending on the magnitude and sign of the Higgs portal coupling $a_2$ and the triplet mass. Correlating this observation with the SM-like Higgs boson diphoton branching fraction, as discussed above, could indicate the presence of the real triplet with SFOEWPT-favorable parameters.

\section{Future Work}

It is clear that there exists a rich set of experimental signatures of the Higgs portal in cosmologically-relevant regions of parameter space. While detailed studies have been carried out for a subset of these signatures, substantial additional work is called for. In Table \ref{tab:signatures} below, we summarize the landscape of completed and prospective studies. From both this discussion and Table  \ref{tab:signatures}  we see that there exist considerable opportunities for both new experimental searches and theoretical studies. As the theoretical work develops, we will periodically up-date this document to reflect new results. We also point to a companion document that summarizes the outcome of a recent ACFI workshop \lq\lq Probing the Electroweak Phase Transition with a 100 TeV pp Collider" that was held at the University of Massachusetts Amherst in September 2015.

\begin{table}[hb]
\small
\begin{centering}
\begin{tabular}{|c|lc|c|c|c}
\hline\hline
Scenario	&  Conditions  & Collider & Signatures & Ref. \\
\hline\hline
Real Singlet		 & EWPT & LHC  & Signal reduction & \cite{Profumo:2007wc,Barger:2007im,Profumo:2014opa}\\ 
&& ILC, FCC-ee,   & Signal reduction & \cite{Profumo:2007wc,Profumo:2014opa}\\
&& CEPC, SppC/FCC-hh &&\\
\hline
&EWPT&LHC & Modified self-coupling & \cite{Profumo:2007wc,Noble:2007kk,Katz:2014bha,Profumo:2014opa}\\
&& ILC, TLEP & Modified self-coupling & \cite{Profumo:2014opa}\\
&& CEPC, SppC/FCC-hh &&\\
\hline
&EWPT&LHC& $h_2\to h_1 h_1\to b{\bar b}\tau^+\tau^-$ & \cite{No:2013wsa}\\
&&& $h_2\to h_1 h_1\to b{\bar b}\gamma\gamma$ & \cite{Chen:2014ask,Barger:2014taa}\\
&&& $h_2\to h_1 h_1\to b{\bar b} b{\bar b}$ & \cite{Chen:2014ask} \\
&&& $h_2\to h_1 h_1\to b{\bar b}VV$ & In prog  \\
&&& $h_1\to h_2 h_2\to XY$ & In prog \\
\hline
Real Singlet		 & DM & LHC  & Invisible decays & \cite{Barger:2007im,He:2013suk,Curtin:2014jma}\\ 
Real Singlet & DM$^\ast$ \& EWPT & SppC/FCC-hh & VBF $h_1^\ast\to$ invis & \cite{Curtin:2014jma}\\ 
&&& invisible decays &\\
&&& modified self-coupling &\\
 & & ILC/FCC-ee/CEPC & modified self-coupling &\cite{Curtin:2014jma}\\
 &&& Signal reduction & \\

\hline\hline
Real Triplet & EWPT \& DM$^\ast$ & LHC & Disappearing charged  & \cite{FileviezPerez:2008bj}\\
& & & particle tracks & \\
&&& Reduced $\Gamma(h\to\gamma\gamma)$ &\cite{Patel:2012pi}\\
\hline\hline
Real Triplet & EWPT only & LHC & Reduced $\Gamma(h\to\gamma\gamma)$& \cite{Patel:2012pi}\\
&&& $H^+ H_2^0\to b{\bar b}\tau\nu$ & \cite{FileviezPerez:2008bj}\\
\hline
&EWPT only&& $H^+ H_2^0\to H_2^0 H_2^0\pi^+$ & \\
&&& $H^+ H^-\to H_2^0 H_2^0\pi^+\pi^-$ & \\
&&& with & \\
&&& $\qquad\qquad H_2^0 H_2^0 \to b{\bar b}\gamma\gamma$ & \cite{FileviezPerez:2008bj}\\
&&& $\qquad\qquad H_2^0 H_2^0 \to b{\bar b}\tau^
+\tau^-$ & \\
\hline\hline
2HDM & EWPT & LHC & $A^0\to Z H^0\to \ell\ell b{\bar b}$ & \cite{Dorsch:2014qja}\\
&&& $A^0\to Z H^0\to \ell\ell W^+ W^-$ &\cite{Dorsch:2014qja} \\
\hline\hline
\end{tabular}
\caption{Higgs portal scenarios signatures.  For each scenario, the cosmological relevance is indicated in the second column. The third column gives the collider for which the signature indicated in the fourth column has been analyzed. The final column gives appropriate references. For exotic Higgs boson decays, see Table \ref{tab:decay_exo}. $^\ast$ For this case, we do not assume saturation of the DM relic density.\label{tab:signatures}}
\end{centering}
\end{table}

\newpage

\section{Acknowledgements }
We thank Xiaochuan Lu for providing Figure \ref{fig:heavysinglet} and Michelangelo Mangano for providing detailed comments on this manuscript. 
SL is supported through FWO Vlaanderen Odysseus II grant G.0C39.13N. X-G He was supported in part by the MOE Academic Excellent Program (Grant No.~102R891505) and MOST of R.O.C. 
(Grant No.~MOST104-2112-M-002-015-MY3), and in part by NSFC (Grant Nos.~11175115 and 11575111) and Shanghai 
Science and Technology Commission (Grant No.~11DZ2260700) of P.R.C. The work of A. Kotwal was supported by the Fermi National Accelerator Laboratory and by a Department of Energy grant to Duke University. Fermilab is operated by Fermi Research Alliance, LLC, under Contract No. DE-AC02-07CH11359 with the United States Department of Energy. J. Kozaczuk is supported by the National Sciences and Engineering Research Council of Canada (NSERC). J.M.N. is supported by the People Programme (Marie Curie Actions) of the European Union Seventh 
Framework Programme (FP7/2007-2013) under REA grant agreement PIEF-GA-2013-625809. MJRM and PW were supported in part under U.S. Department of Energy contract DE-SC0011095. 




\newpage
\vspace*{5mm}

\bibliographystyle{JHEP.bst}
\bibliography{HiggsPortalRefs}

\end{document}